\documentclass[10pt,superscriptaddress,aps,prapplied,twocolumn,nofootinbib]{revtex4-2}
\usepackage{amsmath,amssymb,amsthm}
\usepackage{multirow}
\usepackage{tabularx}
\usepackage[breaklinks=true,colorlinks,citecolor=blue,linkcolor=red,urlcolor=blue]{hyperref}
\usepackage[pdftex]{graphicx}
\usepackage{times,txfonts}
\usepackage{orcidlink}
\usepackage{braket}
\usepackage{color}
\usepackage{natbib}
\usepackage{url}
\usepackage{xurl}
\usepackage{makecell}
\usepackage{bm}
\usepackage{amsmath,blkarray}
\usepackage{mathtools}
\usepackage{makecell}
\usepackage[normalem]{ulem}
\usepackage{latexsym}
\usepackage{tabularx, booktabs}
\usepackage{graphics,epstopdf}
\usepackage{graphicx}
\usepackage[linesnumbered,lined,ruled,vlined]{algorithm2e}
\usepackage{booktabs}
\usepackage{amsfonts}
\usepackage{tikz}
\usepackage{graphicx}
\usepackage{subfigure} 

\usepackage{graphicx}
\usepackage{dcolumn}
\usepackage{bm}

\newcommand{\beq}{\begin{equation}}
\newcommand{\eeq}{\end{equation}}
\newcommand{\beqa}{\begin{eqnarray}}
\newcommand{\eeqa}{\end{eqnarray}}

\usepackage{xcolor}

\begin{document}

\preprint{APS/123-QED}

\title{Shortcuts for adiabatic and variational algorithms in molecular simulation}

\author{Juli\'an Ferreiro-V\'elez$^{\orcidlink{0000-0003-4864-7623}}$}
\email{julian.ferreiro@tecnalia.com}
\affiliation{Department of Physical Chemistry, University of the Basque Country UPV/EHU, Apartado 644, 48080 Bilbao, Spain}
\affiliation{TECNALIA, Basque Research and Technology Alliance (BRTA), 48160 Derio, Spain}
\affiliation{EHU Quantum Center, University of the Basque Country UPV/EHU, 48940 Leioa, Spain}

\author{I\~naki Iriarte-Zendoia$^{\orcidlink{0009-0009-8314-3309}}$}
\affiliation{Department of Physical Chemistry, University of the Basque Country UPV/EHU, Apartado 644, 48080 Bilbao, Spain}	
\affiliation{EHU Quantum Center, University of the Basque Country UPV/EHU, 48940 Leioa, Spain}

\author{Yue Ban$^{\orcidlink{0000-0003-1764-4470}}$}
\email{yue.ban@csic.es}
\affiliation{Quantum Advanced Research Center (QuARC), CSIC, 28049,  Madrid, Spain}
\affiliation{Instituto de Ciencia de Materiales de Madrid (ICMM), CSIC, 28049,  Madrid, Spain}

\author{Xi Chen$^{\orcidlink{0000-0003-4221-4288}}$}
\email{xi.chen@csic.es}
\affiliation{Quantum Advanced Research Center (QuARC), CSIC, 28049,  Madrid, Spain}
\affiliation{Instituto de Ciencia de Materiales de Madrid (ICMM), CSIC, 28049,  Madrid, Spain}

\date{\today}

\begin{abstract}
Quantum algorithms offer a promising route toward computational advantage, but current implementations remain constrained by limited coherence and gate errors, making circuit complexity and noise sensitivity critical considerations. Here, we develop a shortcuts-to-adiabaticity strategy for molecular ground-state preparation and apply it to both digitized adiabatic quantum computing and the variational quantum eigensolver. In the adiabatic framework, we construct an approximate counterdiabatic (CD) Hamiltonian using a nested-commutator expansion. For the molecular systems considered, the CD correction accelerates state preparation and improves energy convergence in the fast-evolution regime, enabling a target accuracy to be reached with a lower implemented-operator count. In the variational setting, we introduce the adiabatic gauge ansatz ($\mbox{AGA}$), obtained by promoting the Pauli-string components generated by the approximate adiabatic gauge potential (AGP) to independent variational generators.
We further propose a reduced variant, $\mbox{AGAR}$, that retains only Pauli generators with weight not larger than two, substantially reducing the ansatz size while maintaining competitive accuracy over the regimes considered. Benchmarks on $\mathrm{LiH}$ and $\mathrm{BeH}_2$ show that the proposed ans\"atze achieve accuracies competitive with established approaches such as UCCSD. Experiments on the \texttt{IBM\_basquecountry} superconducting quantum processor, together with simulations under depolarizing noise, show that the reduced CD-inspired ansatz yields smaller energy discrepancies than UCCSD under the noise conditions considered. Overall, our results demonstrate how AGP-inspired operator structures can be used to balance accuracy and implementation cost in molecular simulations on current noise-limited quantum hardware.

\end{abstract}

\maketitle


\section{Introduction}

In recent decades, quantum technologies have made remarkable progress in the engineering and coherent control of quantum systems \cite{Dowling}. These advances have enabled quantum protocols that exploit fundamental quantum features for computing \cite{Cirac_computing}, communication \cite{communication1,communication2}, and sensing \cite{quantum_sensing,sensing_nv}. Since the inception of quantum computing in the 1980s, the simulation of quantum many-body systems has been regarded as a central benchmark and a primary driving motivation for the field \cite{feynman1982simulating}. By harnessing entanglement and the exponential scaling of Hilbert space, quantum processors aim to tackle problems that are intrinsically intractable for classical simulation.

Within this context, chemical simulation is widely viewed as one of the most promising near-term applications of quantum computing. Quantum chemistry seeks to describe molecular and material properties from first principles \cite{qchem,RevModPhys.79.291}. While classical electronic-structure methods can achieve high accuracy, their computational cost often scales exponentially for strongly correlated systems. Quantum computers, in contrast, can emulate many-electron dynamics directly through efficient state encoding and controlled time evolution. Several quantum algorithms have been developed for molecular simulation, including (i) quantum phase estimation (QPE) \cite{kitaev1995quantum}, which uses the quantum Fourier transform to extract eigenvalues and eigenstates \cite{LloydQPE,scienceAspuru}; (ii) adiabatic quantum computing (AQC) \cite{farhi2000quantum,Hauke_2020}, which prepares the molecular ground state via a slow interpolation between an easily prepared initial Hamiltonian and the target Hamiltonian; and (iii) the variational quantum eigensolver (VQE) \cite{Peruzzo2014,Rev_QCHEM}, a hybrid quantum--classical method that minimizes the energy expectation value over parameterized quantum circuits.

\begin{figure*}
    \centering
    \includegraphics[width=\textwidth]{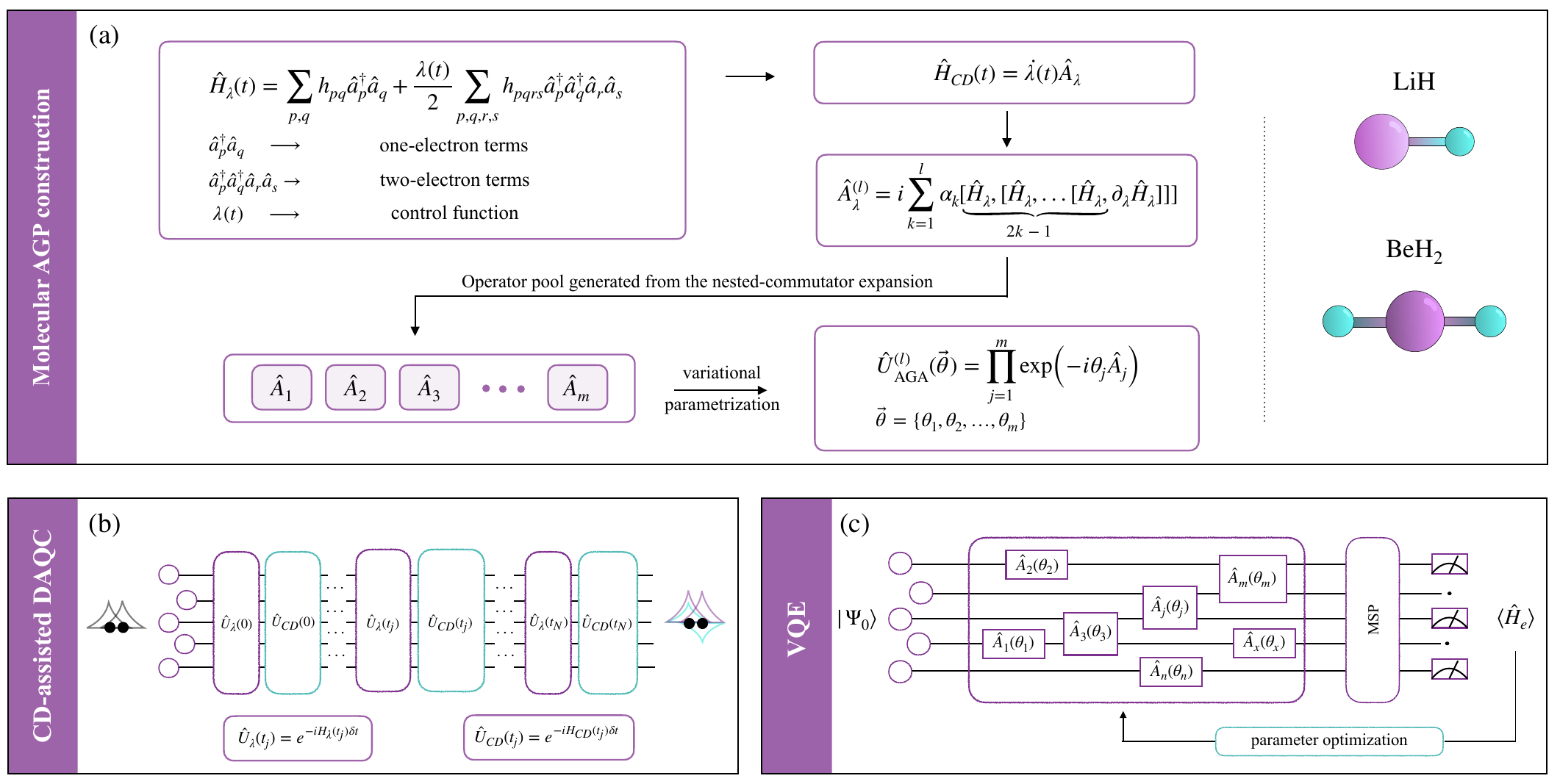}
    \caption{Schematic representation of the CD protocol, rooted in STA, for molecular ground-state preparation. (a) Starting from the molecular electronic Hamiltonian $(\hat{H}_\lambda)$, the approximate AGP is constructed through a truncated nested-commutator expansion. The index $k$  labels the term in the nested-commutator expansion, while $l$ denotes the truncation order. The resulting operators define the AGP operator space. The $\mbox{AGA}$ ansatz is constructed by assigning independent variational parameters $\theta_j$ to the resulting Pauli-string generators $\{\hat A_j\}_{j=1}^{m}$. (b) The Adiabatic block depicts the accelerated evolution, where the system is driven from a noninteracting reference to the fully interacting molecular configuration by incorporating the CD Hamiltonian, enabling faster convergence to the ground state. (c) The VQE block presents the variational implementation based on the $\mbox{AGA}$, including the parameterized circuit, measurement state preparation, and classical optimization loop.}
    \label{fig: AGA-FIG}
\end{figure*}

Despite substantial experimental progress, current quantum processors remain limited by decoherence, gate errors, and restricted connectivity \cite{Preskill2018quantum}. These hardware constraints limit the depth and accuracy of executable quantum circuits and hinder the practical realization of fault-tolerant algorithms such as QPE, Shor’s \cite{shor1994algorithms}, or Grover’s algorithms \cite{grover1996fast}. For current noise-limited quantum hardware, adiabatic and variational approaches provide alternatives that avoid some of the requirements of fully fault-tolerant algorithms, although their implementation remains constrained by circuit depth and control complexity~\cite{RMPNISQ22,RevModPhys.95.045005}. Adiabatic algorithms for molecular simulation have received comparatively limited attention \cite{1AdiabMol,mdlAdiabMol}, in part because implementing time-dependent molecular Hamiltonians typically requires deep circuits with many gates. Although recent work has improved their scalability \cite{DwaveAdiab,CiracAdiab}, extending them to larger and more complex molecules remains challenging. Meanwhile, VQE has been extensively explored and has demonstrated strong performance for small molecules \cite{IonVQE,super-VQE,VQEprx2018,Jian-WeiPan}. However, its success depends critically on the choice of ansatz. Compact yet expressive ans\"atze are essential for reducing circuit depth, limiting noise accumulation, and avoiding overly redundant parameterizations that may hinder optimization \cite{BP1,BP2,BP3}.
Existing approaches include hardware-efficient circuits \cite{Hrdweff,Hrdweff2}, chemically inspired ans\"atze such as UCCSD \cite{UCCSD,KupCCGSD}, and Hamiltonian-informed constructions \cite{Haminsp1}, as well as adaptive strategies such as ADAPT-VQE \cite{grimsley2019}.

Motivated by these challenges, we turn to shortcuts to adiabaticity (STA) \cite{RevModPhys.91.045001}, and in particular CD driving, as a strategy to enhance both adiabatic and variational algorithms for molecular ground-state preparation. CD driving \cite{Berry_2009,Demirplak_Rice_2003} introduces an auxiliary Hamiltonian that suppresses diabatic transitions and enforces evolution along the instantaneous eigenstate manifold, enabling faster state preparation. Although exact CD driving can be formally derived \cite{XiPRL,AdolfoPRL,PhysRevXCDAdolfo,LMGmodelPFL}, it requires knowledge of the instantaneous eigenstates and typically generates highly nonlocal many-body interactions, making it intractable for generic systems and challenging to implement on current hardware. 

To overcome these limitations, a variety of approximate and local CD constructions have been developed, including variational approaches \cite{Claeys,CD-variational}, learning-based strategies \cite{RLCD,ferrer2024physics}, and CD optimized local driving (COLD) \cite{COLDPRXQuantum24}. These developments have enabled improved control of complex many-body systems while maintaining experimentally feasible locality. Moreover, their structured form has facilitated the integration of CD concepts into quantum algorithms, including digitized adiabatic quantum computing (DAQC) \cite{DIG-STA,PengPRApplied,LucignanoPRR,KeeverPRXQuantum24} and optimized annealing protocols \cite{TakahashiPRA,JPSJ,Hartmann_2019,2parameterPRR}. CD constructions have also been extended to variational quantum algorithms, where auxiliary AGP-inspired operators are incorporated into variational circuits. Applications include quantum optimization~\cite{QAOAlove,PranavQAOA,BenchmarkingPRR24} and electronic-structure ground-state preparation~\cite{YinPRApplied21,ferrer2024physics,tang2024exploring,PhysRevApplied.21.034033}, where AGP-derived operator structures provide physically motivated alternatives for variational circuit design. More recently, adaptive strategies have combined CD-inspired operator pools with iterative ansatz construction, providing a complementary route toward compact variational circuits \cite{tancara2026counterdiabatic}.

In this work, we apply approximate CD driving constructed from a nested-commutator expansion \cite{Claeys,CD-variational} to accelerate digitized adiabatic algorithms for molecular ground-state preparation, as schematically illustrated in Fig.~\ref{fig: AGA-FIG}.  The CD-assisted evolution improves convergence and, for the molecular systems considered here, reduces the implementation resources required to reach a target accuracy in the fast-evolution regime. We then translate the same gauge-potential structure into the variational setting by introducing a CD-inspired ansatz, the {adiabatic gauge ansatz} ($\mbox{AGA}$) [Fig.~\ref{fig: AGA-FIG}(c)].
The $\mbox{AGA}$ and its reduced variant leverage dynamical information
encoded in the approximate CD construction to build physically motivated
variational circuits, providing a tunable trade-off between expressivity
and implementation cost.
Unlike explicitly time-dependent CD protocols, the variational approach absorbs the optimization of the gauge-potential coefficients into the classical feedback loop, providing a practical route to exploit CD-inspired structure within VQE.

The remainder of this article is organized as follows. Sec.~\ref{STA} introduces the molecular Hamiltonian and the CD formalism within the DAQC framework. Sec.~\ref{CD-adiabatic} presents CD-assisted digitized adiabatic ground-state preparation. Sec.~\ref{VQE} describes the $\mbox{AGA}$ construction and its reduced variant, $\mbox{AGAR}$, within the VQE framework and benchmarks their performance against established ansätze. Sec.~\ref{discussion} presents hardware implementations on the \texttt{IBM\_basquecountry} quantum processor and complementary noisy simulations. Finally, Sec.~\ref{Conclusion} concludes and provides an outlook.

\section{\label{STA} PRELIMINARIES}

\subsection{Molecular Hamiltonian and Qubit Encoding}

Quantum chemistry aims to determine molecular properties from first principles by solving the electronic Schr\"odinger equation. Since exact solutions are limited to simple systems, practical calculations rely on controlled approximations and efficient representations of the molecular Hamiltonian. Within the nonrelativistic approximation, the molecular Hamiltonian is written as \cite{qchem,RevModPhys.79.291,Rev_QCHEM}
\begin{equation}
\begin{aligned}
\hat{H}_{\text{mol}} = &- \sum_{i=1}^{N} \frac{\hbar^2}{2m_e}\nabla_i^2 
- \sum_{I=1}^{K} \frac{\hbar^2}{2M_I}\nabla_I^2 
- \sum_{i=1}^{N}\sum_{I=1}^{K} \frac{e^2 Z_I}{4\pi \epsilon_0 |\vec{r}_i-\vec{R}_I|} \\
&+ \frac{1}{2}\sum_{i\neq j}^{N}\frac{e^2}{4\pi \epsilon_0 |\vec{r}_i-\vec{r}_j|} 
+ \frac{1}{2}\sum_{I\neq J}^{K}\frac{e^2 Z_I Z_J}{4\pi \epsilon_0 |\vec{R}_I-\vec{R}_J|}.
\end{aligned}
\label{eq:full_molecular_hamiltonian_app}
\end{equation}
Here $N$ and $K$ denote the number of electrons and nuclei, $m_e$ and $M_I$ are the electron and nuclear masses, $\vec{r}_i$ and $\vec{R}_I$ represent their spatial coordinates, and $Z_I$ is the atomic number of nucleus $I$. The first two terms describe the kinetic energies of electrons and nuclei, while the remaining terms account for Coulomb interactions.

The Born-Oppenheimer approximation exploits the large mass difference between nuclei and electrons, allowing the nuclei to be treated as fixed classical point charges. 
Under this approximation, the molecular wavefunction can be approximated as $\Psi(\vec{r},\vec{R}) \simeq \psi_e(\vec{r};\vec{R})\,\chi_n(\vec{R})$, where $\psi_e$ and $\chi_n$ describe the electronic and nuclear components, respectively.
By treating the nuclei as fixed sources of electrostatic potential,  the electronic Hamiltonian (in atomic units $\hbar = m_e = e = 4\pi\epsilon_0 = 1$) becomes
\begin{equation}
\hat{H}_{e}=
-\sum_{i=1}^{N}\frac{1}{2}\nabla_i^2
-\sum_{i=1}^{N}\sum_{I=1}^{K}\frac{Z_I}{r_{iI}}
+\sum_{i<j}^{N}\frac{1}{r_{ij}},
\label{eq:electronic_hamiltonian}
\end{equation}
where $r_{iI}=|\vec r_i-\vec R_I|$ and
$r_{ij}=|\vec r_i-\vec r_j|$. The nuclear-nuclear repulsion term can be treated as a geometry-dependent energy shift and is therefore not included in the electronic Hamiltonian.

To facilitate both analytical treatments and numerical simulations, we  express the electronic Hamiltonian in the second-quantization form:
\begin{equation}
\hat{H}_e=
\sum_{p,q}h_{pq}\hat{a}_p^\dagger\hat{a}_q+
\frac{1}{2}\sum_{p,q,r,s}h_{pqrs}\hat{a}_p^\dagger\hat{a}_q^\dagger\hat{a}_r\hat{a}_s.
\label{eq:second_quantized_H}
\end{equation}
Here, $\hat{a}_p^\dagger$ and $\hat{a}_p$ are fermionic creation and annihilation operators, respectively, and the electronic Hamiltonian is represented in a basis of antisymmetric spin orbitals $\{\phi_p\}$. The coefficients $h_{pq}$ and $h_{pqrs}$ denote the one- and two-electron integrals, respectively, which encode the one- and two-electron contributions of the molecular Hamiltonian \cite{Rev_QCHEM,McClean_2020}. The electronic states are represented in the occupation-number basis, where each spin orbital $\phi_p$ has occupation $f_p\in\{0,1\}$.

To implement Eq.~(\ref{eq:second_quantized_H}) on a gate-based quantum computer, fermionic operators 
are mapped to qubit operators through transformations that express them as combinations of tensor products of Pauli matrices, $\{\hat{a}_p^\dagger,\hat{a}_p\}\rightarrow\{\hat{X},\hat{Y},\hat{Z}\}$. Common mappings include the Jordan-Wigner (JW) transformation~\cite{JW}, the parity mapping~\cite{bravyik}, and the Bravyi-Kitaev (BK) transformation~\cite{bravyi2002fermionic}. In this work, we adopt the BK transformation, which combines occupation and parity information and can reduce the support of fermionic operators relative to the JW mapping, potentially lowering the implementation cost.

\subsection{Counterdiabatic Driving}

Throughout this work, we develop CD strategies to accelerate molecular ground-state preparation on quantum devices. As a starting point, we consider an adiabatic interpolation of the molecular Hamiltonian in Eq.~(\ref{eq:second_quantized_H}),
\begin{equation}
\hat{H}_\lambda(t)=\sum_{p,q} h_{pq}\,\hat{a}_p^\dagger \hat{a}_q
+\frac{\lambda(t)}{2}\sum_{p,q,r,s} h_{pqrs}\,\hat{a}_p^\dagger\hat{a}_q^\dagger\hat{a}_r \hat{a}_s,
\label{eqn:second_quantization_adiab}
\end{equation}
where $\lambda(t)\in[0,1]$ controls the strength of the electron--electron interaction. This interpolation connects the non-interacting reference Hamiltonian to the fully interacting molecular Hamiltonian. As $\lambda(t)$ increases, electronic correlations are progressively introduced, driving the system toward the correlated ground state.

In practice, the complex spectral structure of molecular Hamiltonians and finite evolution times lead to diabatic excitations during the evolution.
To mitigate these effects, we incorporate CD control by augmenting the Hamiltonian as
\begin{equation}
\label{cd_texto}
\hat{H}(t)=\hat{H}_\lambda(t)+\hat{H}_{CD}(t),\qquad
\hat{H}_{CD}(t)=\dot{\lambda}(t)\hat{A}_\lambda,
\end{equation}
where $\hat{A}_\lambda$ is the adiabatic gauge potential (AGP). We adopt a smooth schedule $\lambda(t)=\sin^2(\pi t/2T)$, which ensures vanishing time derivatives at boundaries $t=0$ and $t=T$, where $T$ is the total evolution time. 
The AGP generates infinitesimal unitary transformations along the adiabatic path, 
suppressing diabatic transitions \cite{Jarzynski_2013}. However, computing the exact AGP requires full knowledge of instantaneous eigenstates, making it as difficult as solving the original problem. 

To address this challenge, variational CD approaches construct approximate AGPs within a restricted operator space \cite{CD-variational}. In particular, the AGP can be approximated using a truncated nested-commutator expansion \cite{Claeys},
\begin{equation}
\label{nested}
        \hat{A}_\lambda^{(l)} = i \sum_{k=1}^l \alpha_k \underbrace{[\hat{H}_\lambda,[\hat{H}_\lambda,...[\hat{H}_\lambda,}_{2k-1}\partial_\lambda \hat{H}_\lambda]]],
\end{equation}
where $k=1,\ldots,l$ indexes the terms of the nested-commutator
expansion, and $l$ denotes the truncation order. The $k$-th term
contains $2k-1$ nested commutators.
Increasing $l$ enlarges the AGP operator space and can generate more nonlocal and higher-body terms. These contributions become particularly relevant near gap-closing points, where long-range CD corrections may be required \cite{Adolfo_QPT_ISING_PRL}. Away from such critical regimes, low-order AGP approximations can capture the dominant CD contributions in gapped systems. We therefore focus on $l=1$ as a practical compromise between expressivity and implementation cost.
The coefficients $\vec{\alpha}$ are determined by minimizing the action functional
\begin{equation}
\label{action}
S_l(\vec{\alpha}) = \mathrm{Tr}\!\left[\hat{G}_l(\vec{\alpha})^2\right],\qquad
\hat{G}_l = \partial_\lambda \hat{H}_\lambda - i [\hat{H}_\lambda, \hat{A}_\lambda^{(l)}].
\end{equation}
This procedure yields the variationally optimal approximation of the AGP within the chosen operator manifold. For larger systems, the operator space can be further optimized using more scalable approximation strategies \cite{RLCD,circuitlearning22,COLDPRXQuantum24}.

\section{Speedup of adiabatic state preparation}
\label{CD-adiabatic} 

In this section, we investigate molecular ground-state preparation within the AQC framework, where the solution is obtained by following the ground state of a slowly varying Hamiltonian. In our setting, the Hamiltonian evolution interpolates from an easily
prepared noninteracting reference Hamiltonian, whose ground state $\ket{\Psi_0}$ is represented by a single Slater determinant, to the fully interacting electronic Hamiltonian in Eq.~(\ref{eq:second_quantized_H}). The electron–electron interaction is gradually introduced through a control schedule $\lambda(t)$. The ideal continuous-time evolution is given by
\begin{equation}
\ket{\Psi(T)} \equiv \hat{U}(T)\ket{\Psi_{\rm 0}}
= \mathcal{T}\exp\!\left(-i\int_{0}^{T}\hat{H}_\lambda(t)\,dt\right)\ket{\Psi_{\rm 0}},
\end{equation}
where \(\mathcal{T}\) denotes time ordering.

To implement this evolution on gate-based hardware, we employ DAQC~\cite{barends2016digitized}, which approximates the time-ordered exponential using a first-order Trotter decomposition. After mapping the fermionic Hamiltonian to qubits, the Hamiltonian is expressed as a sum of Pauli strings, $\hat{H}_\lambda(t)=\sum_j \hat{H}_j(t)$, and the evolution operator is approximated as
\begin{equation}
\label{trotterized_evolution}
\mathcal{T}\exp\left(-i\int_{0}^{T}\hat{H}_\lambda(t)\,dt\right)
\approx
\prod_{n=1}^{N} \prod_{j}\exp\!\left[-i\delta t\,\hat{H}_j(n\delta t)\right],
\end{equation}
where $N$ is the number of Trotter steps and $\delta t=T/N$.

\begin{figure}[t]
    \centering
    \includegraphics[width=\columnwidth]{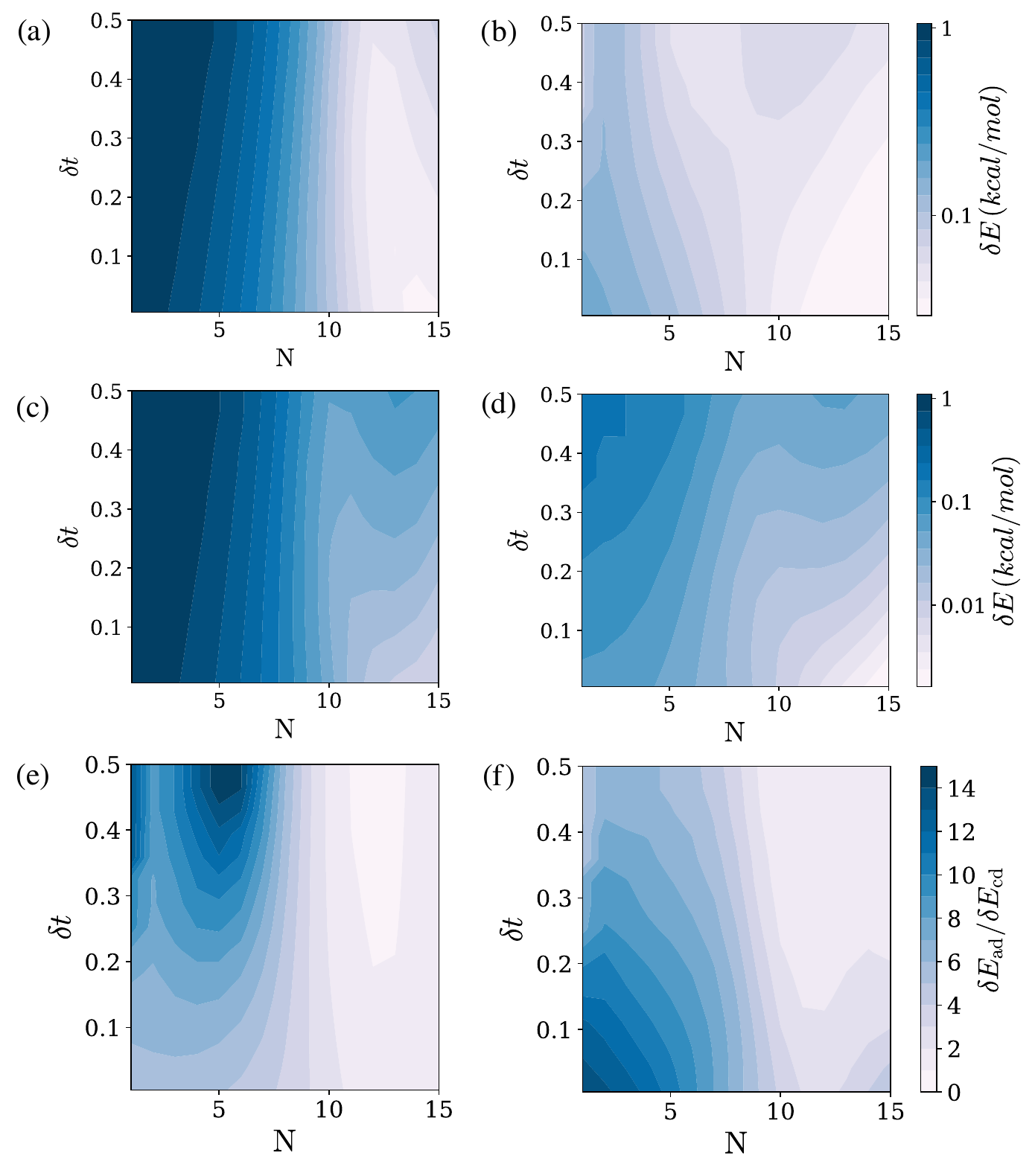}
    \caption{
Ground-state energy discrepancy $\delta E$ [Eq.~(\ref{eqn:trotter_error})] for $\mathrm{LiH}$ [panels (a,b)] and $\mathrm{BeH}_2$ [panels (c,d)] as a function of the Trotter step size $\delta t$ and number of steps $N$. Panels (a,c) and (b,d) show the results without and with the CD correction,
respectively. Panels (e) and (f) show the corresponding improvement ratio for $\mathrm{LiH}$ and $\mathrm{BeH}_2$.}
    \label{fig: adiabatic_evol}
\end{figure}

We now augment the molecular Hamiltonian with a CD correction $\hat{H}_{CD}(t)=\dot{\lambda}(t)\hat{A}_\lambda$ [Eq.~(\ref{cd_texto})], where the AGP $\hat{A}_\lambda$ is approximated using the first-order nested-commutator expansion ($l=1$) in Eq.~(\ref{nested}).  In practice, we construct $\hat{H}_{CD}(t)$ for the $\mathrm{LiH}$ and $\mathrm{BeH}_2$ molecules and incorporate it into the digitized circuits, as illustrated in Fig.~\ref{fig: AGA-FIG}(b). The resulting AGP contains multiple Pauli operators with different supports. The corresponding AGP-derived operator sets used in the variational construction are detailed in Appendix~\ref{appendix: CD Hamiltonians}. Including $\hat{H}_{ CD}(t)$ can improve ground-state preparation at shorter evolution times $T$ by suppressing diabatic transitions.
To quantify this advantage, we compare the energy convergence for different $\delta t$ and $N$, with and without CD correction, for $\mathrm{LiH}$ and $\mathrm{BeH}_2$ (Fig.~\ref{fig: adiabatic_evol}). 

We quantify convergence through the energy error
\begin{equation}
\delta E = E(T;{R}_b)-E_0({R}_b),
\label{eqn:trotter_error}
\end{equation}
where
$E(T;{R}_b)= \langle\Psi(T;{R}_b)|\hat{H}(T)
|\Psi(T;{R}_b)\rangle$
is the final energy obtained from the digitized evolution generated by $\hat{H}(t)=\hat{H}_\lambda(t)+\hat{H}_{CD}(t)$.
At the end of the evolution, \(\dot{\lambda}(T)=0\), so that $\hat{H}_{CD}(T)=0$ and $\hat{H}(T)$ reduces to the target molecular Hamiltonian. Here, ${R}_b$ denotes the interatomic distance and $E_0({R}_b)$ is the exact ground-state energy obtained by diagonalization. In our simulation, the energy is evaluated at fixed bond distances ${R}_b=1.55 \AA$ for the $\rm{LiH}$ and ${R}_b=1.33 \AA$ for the $\rm{BeH}_2$.

Figure~\ref{fig: adiabatic_evol} shows that including $\hat{H}_{ CD}(t)$ improves convergence over a broad range of $\delta t$ and $N$. While the inclusion of $\hat{H}_{CD}(t)$ increases the number of Hamiltonian terms, and hence the implementation cost per Trotter step, it allows for shorter total evolution times. For the molecular systems considered here, this trade-off results in a net reduction of the total implemented-operator count required to reach a given accuracy in the fast-evolution regime; see Appendix~\ref{appendix: CD resources}. Fig.~\ref{fig: adiabatic_evol} further illustrates the interplay between digitization errors and finite-time diabatic errors. For a first-order Trotter decomposition, the leading splitting error arises from the noncommutativity of the Hamiltonian terms within each Trotter slice and scales as
\begin{equation}
\epsilon_{\mathrm{tr}}^{(1)}\sim\mathcal{O}\,\left(T\delta t\,\max_{t\in[0,T]}\sum_{j<j'}\left\|[\hat{H}_j(t),\hat{H}_{j'}(t)]\right\|\right).
\end{equation}
Finite protocol duration introduces an additional diabatic error associated with transitions between instantaneous eigenstates. Writing the evolution in terms of the dimensionless parameter $s=t/T$, the leading adiabatic contribution can be schematically estimated as
\begin{equation}
\epsilon_{\mathrm{ad}}
\sim
\mathcal{O}\,\left(
\frac{1}{T}
\max_{s\in[0,1]}
\frac{\|\partial_s \hat{H}_{\lambda}(s)\|}
{\Delta_{\lambda}(s)^2}
\right),
\end{equation}
where $\Delta_{\lambda}(s)$ denotes the relevant instantaneous energy gap of the reference Hamiltonian $\hat H_{\lambda}(s)$.

By construction, the CD term suppresses finite-time diabatic transitions. Once digitized, however, the additional CD terms also modify the commutator structure of the Hamiltonian and therefore do not necessarily reduce the Trotter error itself. For $\mathrm{LiH}$ [Fig.~\ref{fig: adiabatic_evol}(e)], the CD advantage persists at relatively large $\delta t$ and small $N$, where digitization effects are more pronounced, indicating that the improvement remains robust in a regime with significant Trotter error. For $\mathrm{BeH}_2$ [Fig.~\ref{fig: adiabatic_evol}(f)], the largest improvements occur at short total evolution times, consistent with a stronger role of finite-time diabatic effects. Overall, the net benefit of CD driving reflects a system-dependent competition between the suppression of diabatic transitions and the additional cost of digitizing the CD-assisted Hamiltonian.

\begin{figure*}[htp]
    \centering
\includegraphics[width=\textwidth,height=12cm]{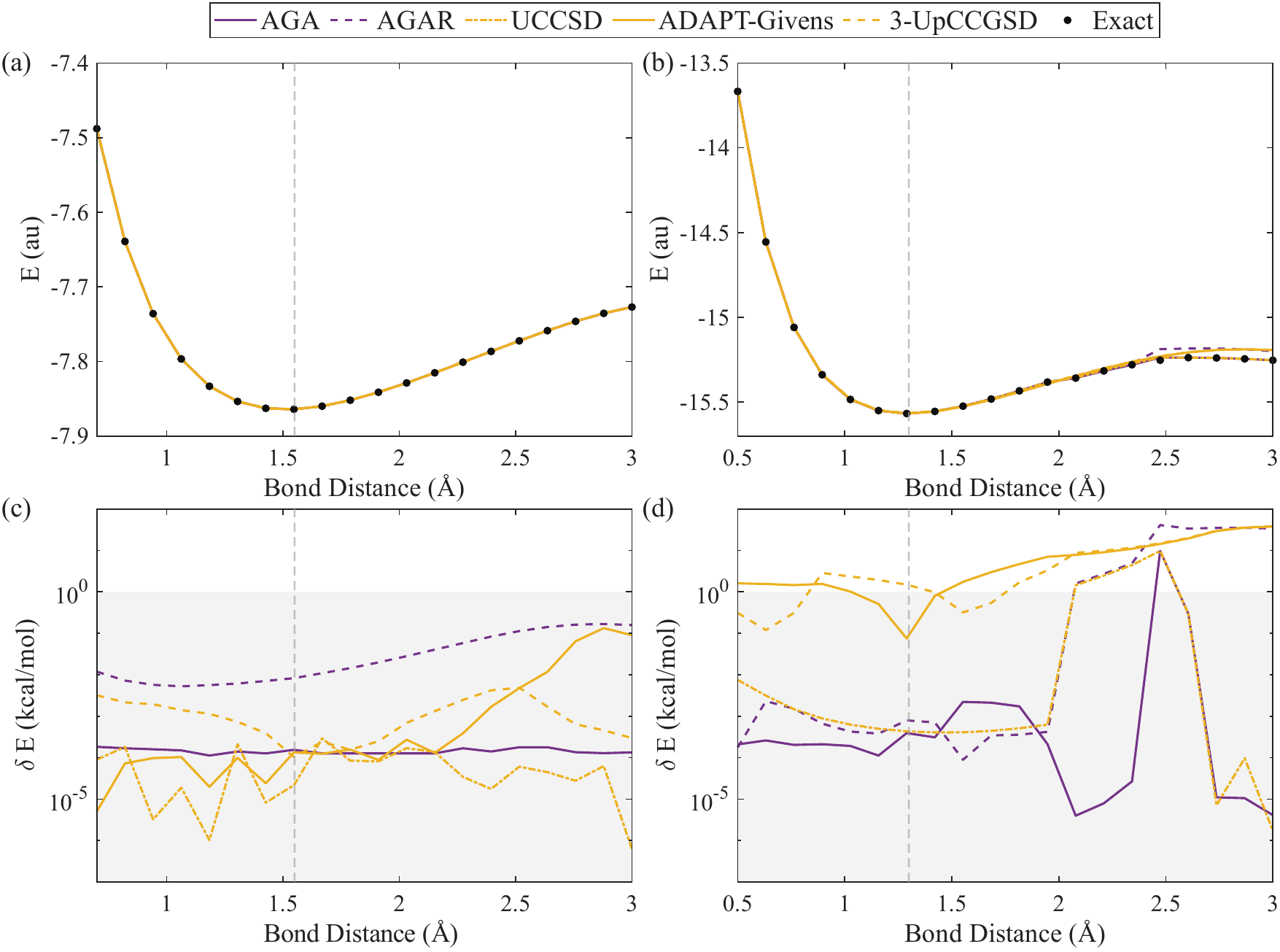}
    \caption{VQE simulations for $\mathrm{LiH}$ (a, c) and $\mathrm{BeH}_2$ (b, d) molecules, comparing several benchmark ans\"atze with $\mbox{AGA}^{(1)}$ and $\mbox{AGAR}^{(1)}$. Panels (a, b) show the ground-state energy as a function of the bond distance, while panels (c, d) display the energy discrepancy between the VQE-optimized energies and exact diagonalization. The vertical dashed line indicates the equilibrium bond distance and the shaded region marks deviations below the chemical-accuracy threshold ($1$ kcal/mol).}
    \label{fig: VQE}
\end{figure*}

Motivated by these results,  we next incorporate the same CD/AGP structure into a variational framework. While AQC and DAQC provide a direct route to state preparation, implementing time-dependent evolutions and CD corrections can be resource intensive as the system size increases. In contrast, VQE transfers part of this complexity to a parameterized ansatz and a classical optimization loop.  The CD-inspired ansatz, namely the $\mbox{AGA}$, therefore provides a variational route to exploit the AGP operator structure without explicitly implementing the full time-dependent CD evolution, allowing a favorable trade-off between accuracy and implementation cost.

\section{ VQE with CD-inspired ansatz}
\label{VQE}

We now turn to the VQE \cite{Peruzzo2014,Rev_QCHEM} for calculating the electronic ground state of molecular systems. VQE is a hybrid quantum--classical algorithm designed to estimate the ground-state energy of many-body Hamiltonians, such as those arising in quantum chemistry. The algorithm employs a parametrized quantum circuit acting on a reference state ($|\Psi_0\rangle$) to generate a variational state

\begin{equation}
\label{eqn: VQE}
|\Psi(\vec{\theta})\rangle
=
\hat U(\vec{\theta})|\Psi_0\rangle,
\end{equation}
where $\hat U(\vec{\theta})$ denotes the parametrized variational circuit and $\vec{\theta}$ is the set of variational parameters. The circuit parameters are optimized classically by minimizing
\begin{equation}
C(\vec{\theta}) = \langle \Psi(\vec{\theta}) | \hat{H} | \Psi(\vec{\theta}) \rangle .
\end{equation}
By the variational principle, minimizing $C(\vec{\theta})$ provides an upper bound to the ground-state energy and drives the variational state toward the ground state within the expressible manifold of the ansatz.
Although the present work focuses on ground-state preparation, extensions of the VQE framework allow access to excited states and spectral properties through modified cost functions and ansatz  constructions \cite{exciteVQE,exciteVQE2}.

The performance of VQE critically depends on the design of the variational  ansatz in $\hat U(\vec{\theta})$ of Eq. (\ref{eqn: VQE}). An effective ansatz should be expressive enough to capture the relevant correlations of the target state while avoiding unnecessary parameters and circuit complexity that can hinder optimization and amplify the effects of {hardware noise} \cite{BP1,BP2,BP3}. 
Inspired by Hamiltonian-variational approaches \cite{hva1,hva2} and CD-inspired circuits \cite{PranavQAOA,DIG-STA}, we introduce an ansatz based on the operator structure generated by the approximate AGP {for the molecule ground-state preparation}. This approach, termed the $\mbox{AGA}$, uses the AGP-derived Pauli operators as a physically motivated variational basis, as illustrated in Fig.~\ref{fig: AGA-FIG}(c). Building upon Eq.~(\ref{nested}), the Pauli-string components generated by the nested-commutator expansion are promoted to independent variational generators. The corresponding $\mbox{AGA}$ circuit is implemented as
\begin{equation}
\label{eqn:reparam}
\hat{U}_{\mathrm{AGA}}^{(l)}(\vec{\theta})
=
\prod_{j=1}^{m}
\exp\!\left(-i\theta_j \hat{A}_j\right),
\end{equation}
where $\{\hat{A}_j\}_{j=1}^{m}$ are the Pauli-string generators obtained
from the nested-commutator expansion truncated at order $l$, $m$ is the
number of generated operators, and $\vec{\theta}=\{\theta_j\}$ are
independent variational parameters. The explicit Pauli-generator sets
used for $\mathrm{LiH}$ and $\mathrm{BeH}_2$ are listed in
Appendix~\ref{appendix: CD Hamiltonians}.

The $\mbox{AGA}$ is a Hamiltonian-informed variational ansatz that incorporates dynamical information from CD driving by using the AGP-generated operators as a physically motivated variational basis.
The approximate AGP identifies physically relevant directions associated with changes along the adiabatic path, motivating their use as variational generators for ground-state preparation. 
Within the VQE framework, the coefficients of the individual AGP-derived Pauli generators are treated as independent variational parameters, rather than being constrained by the coefficients $\{\alpha_k\}$ of the approximate AGP in Eq.~(\ref{nested}) \cite{GeneticCD,RLCD,circuitlearning22,BenchmarkingPRR24}. This reparameterization increases the variational flexibility within a given AGP-generated operator space and can reduce the need for higher-order nested-commutator expansions when the dominant variational directions are already captured at low order. For $\mathrm{LiH}$, increasing the truncation order from $l=1$ to $l=2$ does not lead to a noticeable improvement, indicating that the first-order operator space already captures the dominant variational directions for this system. Nevertheless, for $\mathrm{BeH}_2$ the inclusion of $l=2$ operators {slightly} improves the convergence. This system-dependent trade-off between expressivity and implementation cost is analyzed in Appendix~\ref{appendix: ansätze parametrization}.

For the $\mathrm{LiH}$ and $\mathrm{BeH}_2$ molecules considered here, even the first-order expansion ($l=1$) generates Pauli operators with support on more than two qubits, which can increase the implementation cost on current quantum hardware. To reduce this complexity, we introduce a simplified variant, the {reduced adiabatic gauge ansatz} ($\mbox{AGAR}$), which retains only Pauli generators of weight not larger than two,
i.e., Pauli strings containing at most two non-identity operators.
This restriction reduces the number of variational generators and the corresponding implementation cost, while preserving competitive accuracy over the regimes analyzed below. To benchmark the proposed $\mbox{AGA}$ and $\mbox{AGAR}$ ansätze, we compare them with widely used ans\"atze, including UCCSD \cite{UCCSD}, 3-UpCCGSD \cite{KupCCGSD}, and ADAPT-Givens VQE \cite{grimsley2019,AddaptPenny} (see Appendix~\ref{Appendix_C:}). Starting from the Hartree–Fock reference state $|\Psi_0\rangle = |\Psi_{\rm HF}\rangle$, we evaluate the ground-state energy and its deviation from exact diagonalization for $\mathrm{LiH}$ and $\mathrm{BeH}_2$ across a range of bond distances. Fig.~\ref{fig: VQE} shows the ground-state energy [panels (a, b)] and corresponding errors [panels (c, d)]. Both $\mbox{AGA}^{(1)}$ and $\mbox{AGAR}^{(1)}$ reach chemical accuracy (below $\sim 1$ kcal/mol) over substantial parts of the bond-distance ranges considered. In particular, $\mbox{AGA}^{(1)}$ generally yields smaller energy discrepancies than the benchmark ans\"atze, while the reduced $\mbox{AGAR}^{(1)}$ provides a favorable trade-off between accuracy and ansatz size.

\begin{table}[t]
\centering
\renewcommand{\arraystretch}{1.5}
\setlength{\tabcolsep}{3.5pt}
\begin{tabular}{l|ccccc}
\toprule
& \footnotesize UCCSD & \footnotesize AGA & \footnotesize AGAR & \footnotesize 3-UpCCGSD & \footnotesize ADAPT-Givens \\
\midrule
\midrule

$\mathrm{H_2}$   & 4 (14)   & 2       & 2      & 18  & 2      \\
$\mathrm{LiH}$   & 20 (130) & 56      & 10     & 54  & [13--20] \\
$\mathrm{BeH_2}$ & 77 (574) & [30--488] & [10--12] & 108 & [10--16] \\
\bottomrule
\end{tabular}
\caption{Ansatz sizes used in the VQE simulations. Entries denote the number of variational parameters. For UCCSD, the values in parentheses give the corresponding number of Pauli terms after fermion-to-qubit mapping and decomposition. For $\mbox{AGA}$ and $\mbox{AGAR}$, each variational parameter is associated with one AGP-derived Pauli generator. The $\mathrm{H}_2$ entries are included for reference, following
Refs.~\cite{ferrer2024physics,YinPRApplied21}. }
\label{term_table}
\end{table}

For $\mathrm{BeH}_2$, the performance becomes more challenging at stretched geometries. At shorter bond distances, UCCSD, $\mbox{AGA}^{(1)}$, and $\mbox{AGAR}^{(1)}$ exhibit small energy discrepancies, whereas at larger bond distances the accuracy of all ans\"atze deteriorates to different degrees. The stronger degradation observed for $\mbox{AGAR}^{(1)}$, ADAPT-Givens, and 3-UpCCGSD is consistent with the increased variational difficulty of stretched geometries, where the Hartree--Fock reference generally provides a poorer approximation to the correlated ground state.

We further compare the ans\"atze sizes summarized in Table~\ref{term_table} for the $\mbox{LiH}$ and $\mbox{BeH}_2$ molecules. A larger number of variational parameters increases the classical optimization burden, while the number and structure of the corresponding generators influence the circuit implementation cost after gate decomposition. In the main simulations, we focus on the first-order truncation ($l=1$) as a practical compromise between variational flexibility and implementation cost; the effect of increasing the truncation order to $l=2$ is analyzed explicitly in Appendix~\ref{appendix: ansätze parametrization}.
Ranges are reported in Table~\ref{term_table} whenever the ansatz size varies with bond distance, over $R_b\in[0.5,3]\,\AA$ for $\mathrm{H}_2$, $R_b\in[0.7,3]\,\AA$ for $\mathrm{LiH}$, and $R_b\in[0.5,3]\,\AA$ for $\mathrm{BeH}_2$. For $\mbox{AGA}$ and $\mbox{AGAR}$, such variations arise from their Hamiltonian-based construction: changes in molecular geometry modify the electronic Hamiltonian and, after fermion-to-qubit mapping and symmetry reduction~\cite{bravyi2017tapering}, can lead to different Pauli representations and therefore different AGP-generated operator sets. For ADAPT-Givens, the geometry dependence instead originates from the adaptive construction, which selects a different number of operators from a predefined pool at each bond distance. This adaptive selection can yield more compact problem-dependent ans\"atze, at the cost of repeated operator-selection measurements and additional classical optimization.

For $\mathrm{LiH}$, $\mbox{AGAR}^{(1)}$ substantially reduces the ansatz size relative to the benchmark constructions considered here. The full $\mbox{AGA}^{(1)}$ uses more variational parameters than UCCSD but fewer Pauli generators than the mapped UCCSD representation reported in Table~\ref{term_table}. For $\mathrm{BeH}_2$, the resource requirements are strongly geometry dependent: $\mbox{AGAR}^{(1)}$ remains compact over the bond-distance range considered, whereas the full $\mbox{AGA}^{(1)}$ can grow substantially at stretched geometries as the underlying Hamiltonian-derived operator space increases. Further details on the construction of the ansätze and their geometry dependence are provided in Appendix~\ref{appendix: CD Hamiltonians}.

\begin{figure}[t]
    \centering
    \includegraphics[width=\columnwidth,height=3.8cm]{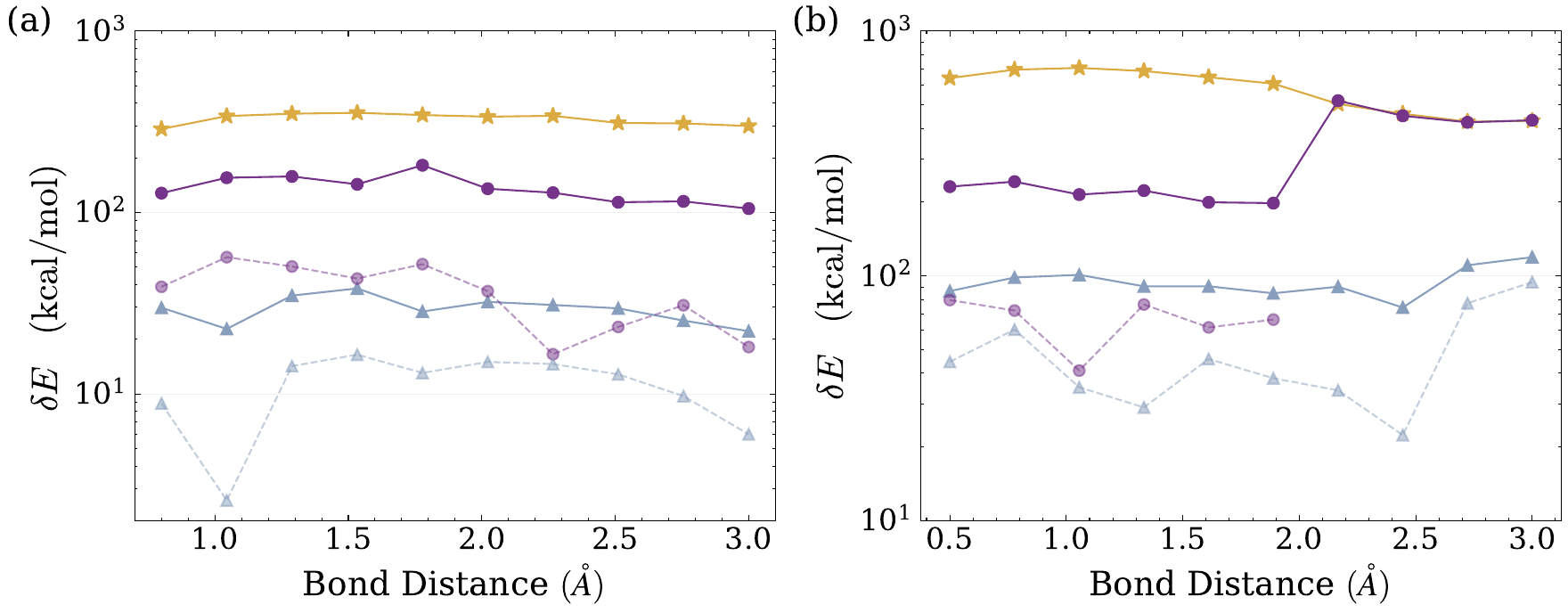}
    \caption{VQE implementation for $\mathrm{LiH}$ (a) and $\mathrm{BeH}_2$ (b) on the \texttt{IBM\_basquecountry} quantum processor. Energy discrepancies $\delta E$ (kcal/mol) obtained with $\mbox{AGA}$ (purple circles), $\mbox{AGAR}$ (blue triangles), and UCCSD (yellow star markers) are shown across different bond distances. Dashed, lighter-shaded lines show the ZNE-mitigated results for AGA and AGAR where circuit size permits use of the extrapolation. 
    }
    \label{fig:real_hw_vqe}
\end{figure}

\section{\label{discussion} Hardware Implementation and Noise Analysis}

We assess the performance of the proposed CD-inspired ans\"atze on current quantum hardware by implementing VQE circuits on the 156-qubit IBM Heron r2 processor \texttt{IBM\_basquecountry}. At the time of execution, the device was characterized by a readout error of $7.14\times10^{-3}$, relaxation and dephasing times of $T_1 = 235.53\,\mu\mathrm{s}$ and $T_2 = 124.64\,\mu\mathrm{s}$, respectively, an SX-gate error of $3.037\times10^{-4}$, and a CZ-gate error of $1.709\times10^{-3}$. The hardware implementation provides a direct benchmark of the proposed ans\"atze in the presence of gate errors, decoherence, and readout imperfections.

Figure \ref{fig:real_hw_vqe} shows the results at different bond distances for the LiH (a) and BeH$_2$ (b) molecules for the UCCSD, AGA, and AGAR ans\"atze. Owing to limited access to quantum hardware, the variational parameters
were first optimized using noiseless simulations and were subsequently
fixed during the hardware runs. For each bond distance, we performed
one hardware evaluation of the corresponding preoptimized parameter set,
rather than reoptimizing the variational parameters on the device. 
Solid lines show results obtained using XY4 dynamical decoupling and
IBM resilience level 1, which provides measurement-error mitigation.
For the AGA and AGAR circuits for which the additional mitigation
overhead remained feasible, we also used resilience level 2, which
additionally enables zero-noise extrapolation (ZNE). We include these results as dashed lines. For UCCSD, and for large bond distances of the BeH$_2$ molecule, the resulting circuit length prevents the use of ZNE, so only resilience level 1 could be applied in these scenarios. Comparing the mitigation-matched results obtained at resilience level 1,
$\mbox{AGAR}$ yields the smallest energy discrepancies among the tested
ans\"atze. 
This behavior is consistent with its reduced operator set and lower implementation cost, which can limit the accumulation of hardware errors. Although chemical accuracy is not reached on the device, the relative performance of the ans\"atze qualitatively follows the expected resource trends.

To separate generic noise effects from device-specific imperfections, we perform complementary noisy VQE simulations using the QIBO framework~\cite{qibo_paper}. We apply a depolarizing channel after each gate operation as a representative stochastic error model,
\begin{equation}
\label{eq:depolarizing_error}
\mathcal{E}_p(\rho)=(1-p)\rho+\frac{p}{4^n-1}\sum_i\mathcal{P}_i \rho \mathcal{P}_i,
\end{equation}
where $\mathcal{P}_i\in\{I,X,Y,Z\}^{\otimes n}\setminus\{I^{\otimes n}\}$, $n$ denotes the number of qubits affected by the channel, and $p$ is the depolarizing strength. This model captures stochastic Pauli errors, including bit- and phase-flip components, and provides a controlled benchmark of the sensitivity of the different ansatz constructions to gate noise.

\begin{figure}[t]
    \centering
    \includegraphics[width=\columnwidth,height=3.8cm]{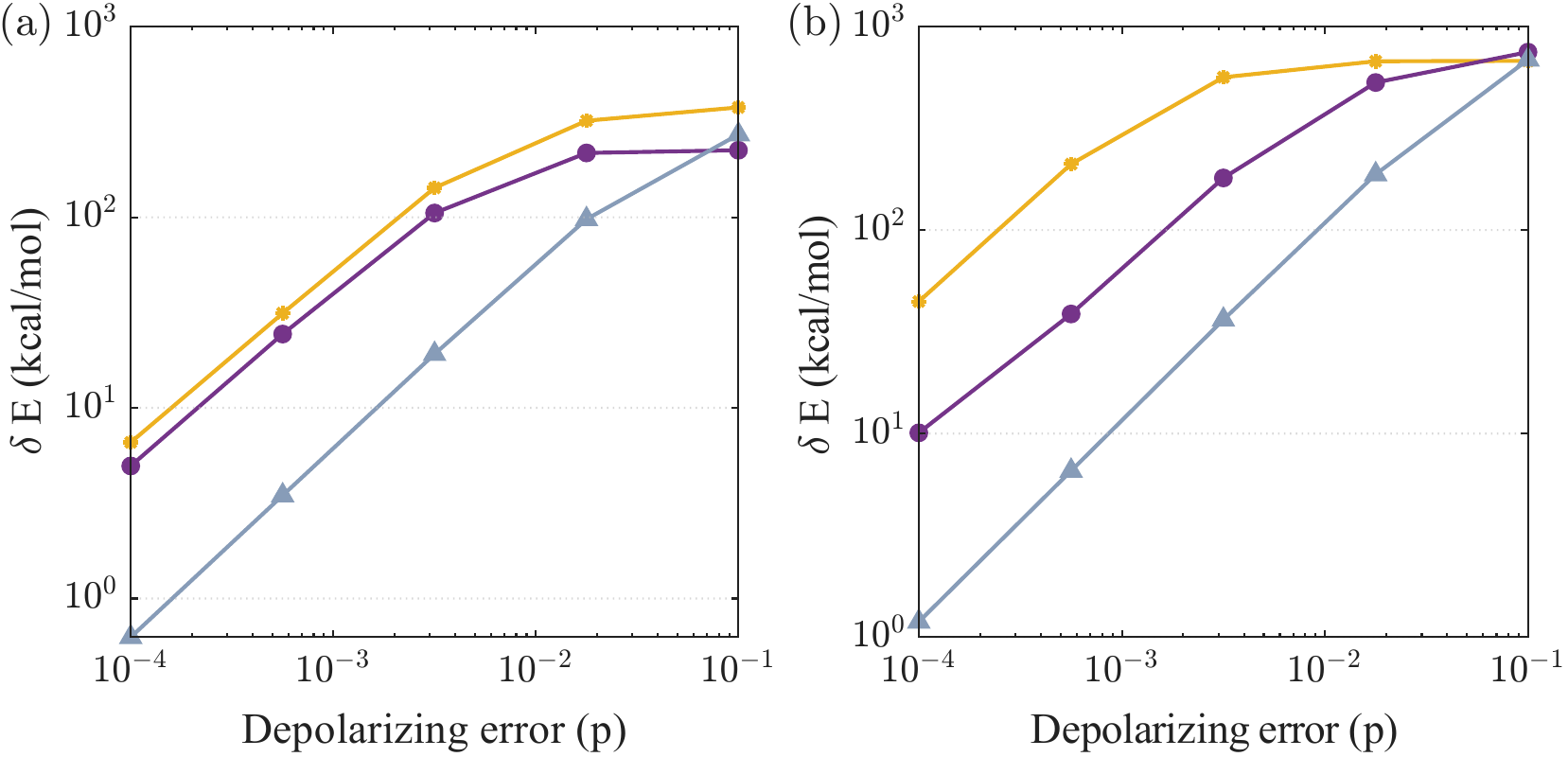}
    \caption{Energy discrepancy $\delta E$ (kcal/mol) between the simulated and exact ground-state energies for $\mathrm{LiH}$ (a) and $\mathrm{BeH}_2$ (b) at their equilibrium bond distances as a function of the depolarizing strength $p$ [Eq.~(\ref{eq:depolarizing_error})]. Results are shown for $\mbox{AGA}$ (purple circles), $\mbox{AGAR}$ (blue triangles), and UCCSD (yellow star markers).}
    \label{fig:depolarizing_error}
\end{figure}

Figure~\ref{fig:depolarizing_error} shows that both $\mbox{AGA}$ and $\mbox{AGAR}$ yield smaller energy discrepancies than UCCSD over the noise range considered. The difference is particularly pronounced for $p\sim10^{-3}-10^{-2}$, where the reduced $\mbox{AGAR}$ ansatz exhibits the smallest error among the three constructions. These results are consistent with the expectation that reducing the number and complexity of implemented generators can limit error accumulation in noisy circuits.

Taken together, the hardware and depolarizing-noise results show that the CD-inspired ans\"atze, and in particular the reduced $\mbox{AGAR}$ construction, can retain higher energy accuracy than UCCSD under the noise conditions considered here. This provides a concrete demonstration of the trade-off between ansatz expressivity and implementation cost on current noise-limited quantum hardware.

\section{\label{Conclusion} Conclusion and Outlook}

In summary, we have investigated CD driving, a central STA technique, for molecular ground-state preparation within both AQC and VQE frameworks. Our results show that CD driving provides a common physical framework for improving finite-time state preparation and for constructing variational ans\"atze from the operator structure of the approximate adiabatic gauge potential. In the molecular examples considered here, these CD-inspired constructions provide favorable trade-offs between accuracy and implementation cost.

In the AQC paradigm, the inclusion of approximate CD terms suppresses finite-time diabatic excitations and improves convergence to the target ground state, particularly in the fast-evolution regime. Although the CD correction increases the implementation cost of each Trotter step, it can reduce the total evolution resources required to reach a given target accuracy. Its net advantage therefore results from a competition between improved adiabatic following and the additional digitization cost associated with the CD-assisted Hamiltonian.

In the variational setting, we introduced the $\mbox{AGA}$, which promotes the Pauli-string components generated by the approximate AGP to independent variational generators. This construction achieves competitive or improved energy accuracy with respect to established ans\"atze such as UCCSD and 3-UpCCGSD for the molecular systems studied here. We further introduced the reduced $\mbox{AGAR}$ ansatz, which retains AGP-derived Pauli generators with weight not larger than two. This reduction substantially decreases the ansatz size while preserving competitive accuracy over the relevant bond-distance regimes.

Beyond noiseless simulations, we evaluated the proposed ans\"atze on the \texttt{IBM\_basquecountry} quantum processor and through controlled depolarizing-noise simulations. Under the noise conditions considered, $\mbox{AGA}$ and particularly $\mbox{AGAR}$ yield smaller energy discrepancies than UCCSD. These results illustrate the practical importance of balancing variational expressivity with implementation cost when designing ans\"atze for current noise-limited quantum hardware.

Overall, our results establish the AGP as a useful bridge between dynamical and variational approaches to molecular ground-state preparation. A natural next step is to combine AGP-derived operator spaces with adaptive selection strategies, in which only the most relevant generators are incorporated during the optimization. Recently proposed adaptive CD VQE approaches provide a complementary realization of this idea \cite{tancara2026counterdiabatic}, trading a more compact problem-dependent circuit for the additional measurement and optimization overhead associated with adaptive operator selection. Further extensions to Krylov-subspace methods~\cite{krylov} and excited-state preparation also represent promising directions.  

From an experimental perspective, CD driving has already been experimentally demonstrated across multiple quantum platforms, including NMR~\cite{Zhou_2024}, photonic systems~\cite{shang2024codesigned,d7bb-ybfh}, and superconducting qubits~\cite{PhysRevApplied.16.034050,Guan_2025}. These developments provide an experimental basis for exploring AGP- and CD-inspired algorithmic constructions on current and future quantum processors.

\section*{Acknowledgment}

This work is partially supported by PID2024-157842OA-I00 and PID2021-126273NB-I00, Projects in the field of Artificial Intelligence 2025 (AIA2025-163435-C44), and HORIZON-CL4-2022-QUANTUM-01-SGA project 101113946 OpenSuperQPlus100 of the EU Flagship on Quantum Technologies. J.F.V. acknowledges support from the UPV/EHU and TECNALIA 2021 PIF contract call, from the Basque Government through the "Plan complementario de comunicación cúantica" (EXP.2022/01341) (A/20220551) and the ELKARTEK program, and project "KUBIT - Kuantikaren Berrikuntzarako Ikasketa Teknologikoa" (KK-2024/00105). I.I.Z. acknowledges support from UPV/EHU Ph.D. Grant No. PIF 23/246. X.C. and Y.B. also acknowledge the Severo Ochoa Centres of Excellence program through Grant CEX2024-001445-S. We acknowledge the use of IBM Quantum services for this work. The views expressed are those of the authors and do not reflect the official policy or position of IBM or the IBM Quantum team.

\section*{Code Availability}
The underlying code for this study will be made available on request from the corresponding author.

\bibliography{bibl}

\appendix

\section{Resources in CD ground-state preparation}
\label{appendix: CD resources}

In the main text, we analyze the impact of the CD term on AQC by comparing the energy convergence of $\mathrm{LiH}$ and $\mathrm{BeH}_2$ with and without the auxiliary CD correction. The CD-assisted protocol improves the convergence in the fast-evolution regime, but this advantage comes at the cost of increasing the number of Hamiltonian terms that must be implemented at each Trotter slice.

\begin{figure}[t]
    \centering
    \includegraphics[width=\linewidth]{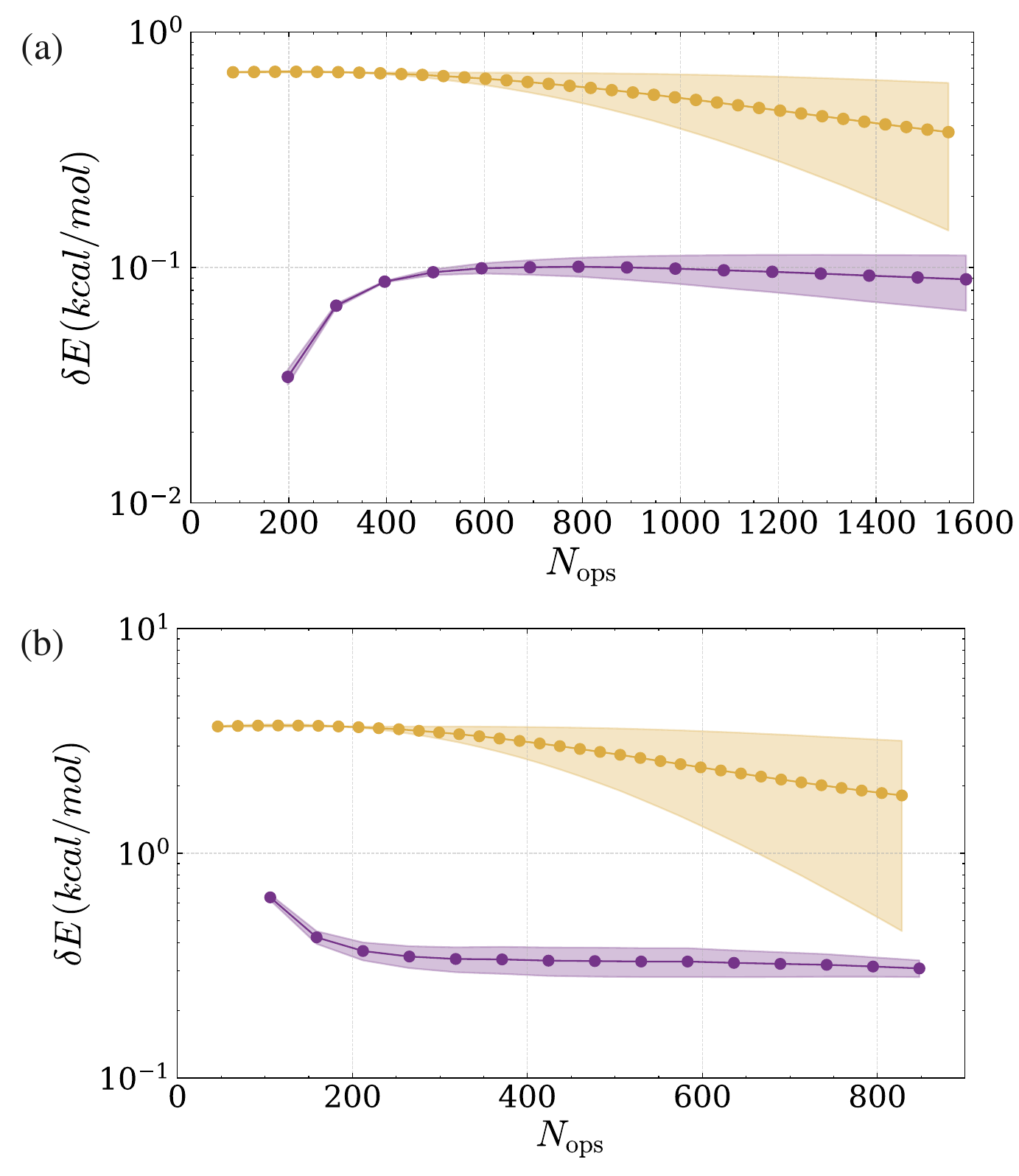}
    \caption{Mean energy discrepancy $\delta E$ as a function of the total number of implemented Hamiltonian terms, used here as a proxy for digital-evolution resources. The CD-assisted evolution (purple) is compared with the unassisted adiabatic evolution (yellow) for $\mathrm{LiH}$ (a) and $\mathrm{BeH}_2$ (b) at their equilibrium bond distances. Solid lines denote the mean over $\delta t\in[0.001,0.3]$, while shaded regions indicate one standard deviation.}
    \label{fig: CD_resources}
\end{figure}

To quantify this trade-off, Fig.~\ref{fig: CD_resources} shows the mean ground-state energy discrepancy as a function of the total number of implemented Hamiltonian terms, $N_{\mathrm{ops}}$. Here, $N_{\mathrm{ops}}$ is used as a proxy for the digital implementation cost rather than as a direct measure of the transpiled circuit depth. The mean is evaluated over Trotter time steps $\delta t\in[0.001,0.3]$, and the shaded region represents one standard deviation.

For both molecules, the CD-assisted evolution reaches smaller energy discrepancies at relatively low $N_{\mathrm{ops}}$, corresponding to the fast-evolution regime. In this region, the suppression of finite-time diabatic excitations compensates for the larger number of operators required in each CD-assisted Trotter slice, allowing a given target accuracy to be reached with a lower overall implemented-operator count. At larger $N_{\mathrm{ops}}$, however, the relative advantage of the CD-assisted protocol is progressively reduced. Because the unassisted Hamiltonian contains fewer terms per Trotter slice, the same operator budget can be distributed over a larger number of evolution steps and therefore over longer total evolution times, $T=N\delta t$. In this regime, the unassisted adiabatic evolution can approach comparable accuracy without the additional CD implementation overhead.

These results show that the resource advantage of the approximate CD protocol is regime dependent. For the molecular systems considered here, CD assistance is most beneficial in the fast-evolution and limited-resource regime, whereas its relative advantage decreases as larger digital-evolution resources become available.

\section{CD-inspired ansatz construction}
\label{appendix: CD Hamiltonians}

In this Appendix, we provide the explicit first-order CD-inspired Pauli generators used to construct the $\mbox{AGA}$ and $\mbox{AGAR}$ ans\"atze for the molecular systems considered in the main text. Starting from the approximate AGP $\hat A_\lambda^{(l)}$ generated by the nested-commutator expansion, its Pauli-string components are denoted by $\{\hat A_j\}_{j=1}^{m}$, consistently with Eq.~(\ref{eqn:reparam}). In the $\mbox{AGA}$, each generator $\hat A_j$ is assigned an independent variational parameter. For $\mbox{AGAR}$, we retain only generators with
Pauli weight not larger than two.

\subsection{Ans\"atze  for $\mathrm{LiH}$}

For $\mathrm{LiH}$, the second-quantized Hamiltonian and its qubit representation contain a large number of terms and are therefore not written explicitly. Instead, we directly list the first-order AGP-derived Pauli generators used in the variational ansatz. The first-order $\mbox{AGA}^{(1)}$ operator set is
\begin{align}
\nonumber
\{\hat A_j\}_{j=1}^{56}=\;& \{IIXY  , IIYX  , IXXY , IXYX, IYII, IYIZ, ZZXY  ,\\ \nonumber
&  IYXX  , IYYY, IYZI , IYZZ , IZXY  , IZYX,\\ \nonumber
& XIIY  , XIYI  , XIYZ , XIZY , XXIY , XXYI  ,\\ \nonumber
& XXYZ , XXZY, XYIX , XYXI  , XYXZ  , XYZX ,\\ \nonumber
& XZIY , XZYI  , XZYZ  , XZZY, YIIX  , YIXI ,  ZZYX,\\ \nonumber
&  YIXZ , YIZX , YXIX  , YXXI , YXXZ, YXZX, \\ \nonumber
&  YYIY  , YYYI  , YYYZ  , YYZY, YZIX   , YZXI  ,\\ \nonumber
&  YZXZ  , YZZX,   ZIXY  , ZIYX , ZXXY , ZXYX  ,\\ \nonumber
& ZYII , ZYIZ  , ZYXX  , ZYYY, ZYZI  , ZYZZ \}. 
\end{align}
Restricting the operator set to Pauli weight not larger than two gives the reduced ansatz
\begin{align}
\nonumber
\{\hat A_j\}_{j=1}^{10}=\;&\{
IIXY  , IIYX  , IYII , IYIZ , IYZI ,\\ \nonumber
& XIIY  , XIYI  , YIIX , YIXI , ZYII\} .
\end{align}
The corresponding $\mbox{AGAR}^{(1)}$ circuit is therefore constructed from ten independent Pauli generators.

\subsection{Ansätze for $\mathrm{BeH}_2$}

For $\mathrm{BeH}_2$, the Pauli structure of the effective qubit Hamiltonian and the resulting AGP-generated operator space can vary with bond distance. We therefore list here the operator sets obtained at a representative equilibrium geometry. The first-order $\mbox{AGA}$ operator set is
\begin{align}
\nonumber
\{\hat A_j\}_{j=1}^{30}=\;& \{IXIY  , IXYI  , IXYZ  , IXZY, \\  \nonumber
& IYIX  , IYXI  , IYXZ  , IYZX ,\\ \nonumber
& XIIY  , XIYI  , XIYZ  , XIZY, \\ \nonumber
& XYII  , XYIZ  , XYZI  , XZIY ,\\ \nonumber
& XZYI  , YIIX  , YIXI  , YIXZ ,\\ \nonumber
& YIZX  , YXII  , YXIZ  , YXZI ,\\ \nonumber
& YZIX  , YZXI  , ZXIY  , ZXYI ,\\
& ZYIX  , ZYXI\}. 
\nonumber
\end{align}

The reduced Pauli-weight $\leq2$ operator set is
\begin{align}
\nonumber
\{\hat A_j\}_{j=1}^{10}=\;& \{
IXIY  , IXYI , IYIX , IYXI, \\ \nonumber
& XIIY  , XIYI  , XYII , YIIX  , YIXI  , YXII\}.
\end{align}
Thus, at this representative geometry, $\mbox{AGAR}^{(1)}$ contains ten independent generators.

\begin{figure}[t]
    \centering
    \includegraphics[width=\linewidth]{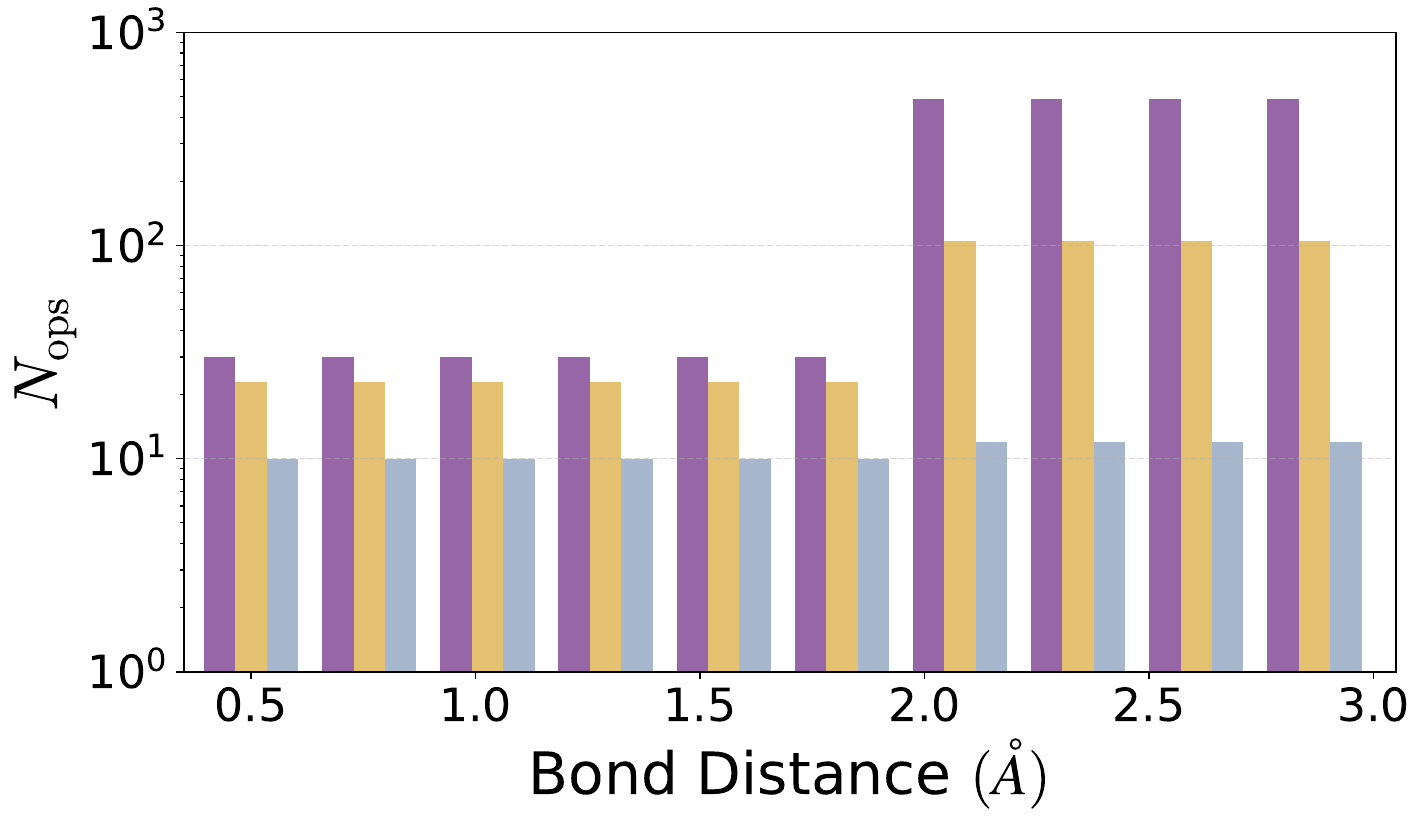}
    \caption{Number of Pauli generators $N_{\mathrm{ops}}$ as a function of bond distance for $\mathrm{BeH}_2$. The operator counts of $\mbox{AGA}^{(1)}$ (purple) and $\mbox{AGAR}^{(1)}$ (blue) are compared with the number of Pauli terms in the corresponding qubit Hamiltonian (yellow).}
    \label{fig: aga_resources}
\end{figure}

Figure~\ref{fig: aga_resources} shows that the size of the $\mathrm{BeH}_2$ operator space increases substantially at stretched geometries. This increase is accompanied by a change in the size and Pauli structure of the effective qubit Hamiltonian. Since the $\mbox{AGA}$ operators are generated directly from the Hamiltonian through the nested-commutator expansion, the enlargement of the mapped Hamiltonian produces a corresponding increase in the number of AGP-derived variational generators.

For $\mathrm{LiH}$, in contrast, the Pauli structure of the effective Hamiltonian remains unchanged over the bond-distance range considered, with the geometry dependence entering primarily through the numerical coefficients. Consequently, the first-order $\mbox{AGA}$ and $\mbox{AGAR}$ operator sets remain unchanged across the geometries analyzed.

\section{AGP reparameterization and higher-order nested expansions}
\label{appendix: ansätze parametrization}

 The construction of variational quantum ans\"atze requires balancing expressivity against implementation cost. Enlarging the operator space, increasing the number of variational parameters, or adding circuit layers can improve the accessible state manifold, but also increases the optimization burden and the cost of implementing the resulting circuit. In this Appendix, we analyze the role of AGP reparameterization and the effect of increasing the nested-commutator truncation order.

As a reference, we consider a layered AGP-inspired variational ansatz constructed from the operator structures generated by the nested-commutator expansion,
\begin{equation}
U_{\mathrm{layer}}^{(l)}(\vec{\theta})=\small{\prod_{p=1}^{N}\prod_{k=1}^{l}\exp\!\left[-i\theta_{pk}\left(i\underbrace{[\hat H_\lambda,[\hat H_\lambda,\ldots[\hat H_\lambda,}_{2k-1}\partial_\lambda \hat H_\lambda]]\right)\right]}.
\end{equation}
Where $N$ is the number of layers, and $l$ the nested-commutator truncation order. Here, the original AGP coefficients $\alpha_k$ are not retained. Instead, the overall amplitude of each nested-commutator contribution is promoted to an independent variational parameter $\theta_{pk}$, while the relative weights of the Pauli-string components within that contribution remain fixed.

\begin{figure}[t]
    \centering
    \includegraphics[width=\linewidth]{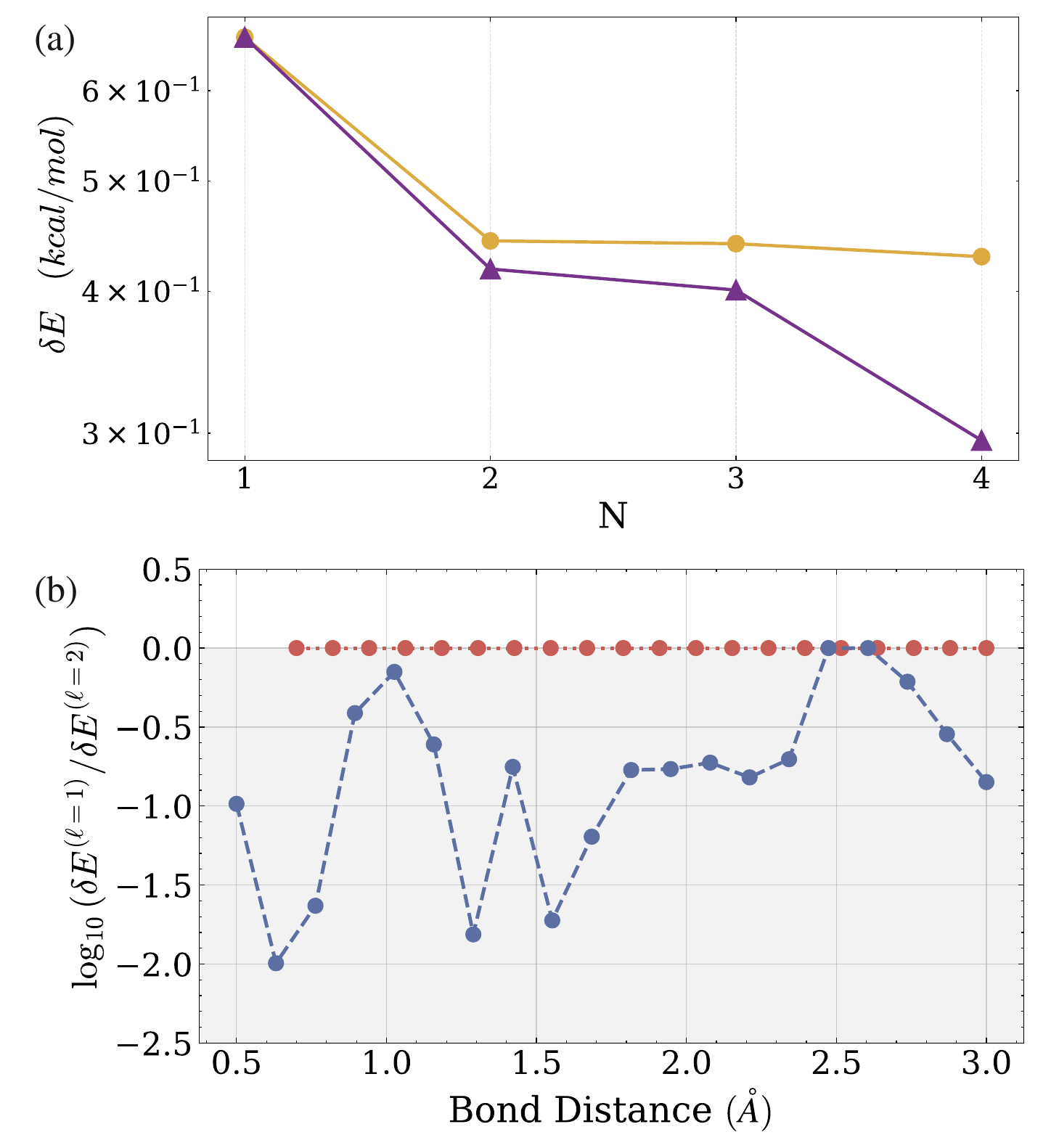}
\caption{Effect of ansatz resources on the variational calculations.
(a) Ground-state energy error for $\mathrm{LiH}$ using an $N$-layer
AGP-inspired ansatz with nested-commutator truncation orders $l=1$
(yellow circles) and $l=2$ (purple triangles) at
$R_b=1.55~\text{\AA}$.
(b) Ground-state energy error ratio for $\mathrm{LiH}$ (red) at
$R_b=1.55~\text{\AA}$ and $\mathrm{BeH}_2$ (blue) at
$R_b=1.33~\text{\AA}$, comparing $\mathrm{AGA}^{(1)}$ and
$\mathrm{AGA}^{(2)}$.}
    \label{fig: vqe_qaoa}
\end{figure}

Figure~\ref{fig: vqe_qaoa}(a) shows the ground-state energy error for $\mathrm{LiH}$ at $R_b=1.55~\text{\AA}$ for $l=1$ and $l=2$ as a function of $N$. Within this layered construction, increasing either $N$ or $l$ improves the variational convergence, at the cost of additional variational parameters and implementation resources.

We next compare this layered construction with the fully reparameterized AGA introduced in the main text. In the AGA, by contrast, the individual Pauli generators $\hat A_j$ are assigned independent variational parameters. This relaxes the fixed relative coefficients within each nested-commutator contribution and thereby increases the variational flexibility of the AGP-derived operator content.

For $\mathrm{LiH}$, the reparameterized $\mbox{AGA}$ reduces the energy discrepancy by approximately three orders of magnitude relative to the layered AGP-inspired construction considered above. Fig.~\ref{fig: vqe_qaoa}(b) further examines the effect of increasing the AGP truncation order within the $\mbox{AGA}$. For $\mathrm{LiH}$, extending the operator space from $l=1$ to $l=2$ does not produce a noticeable improvement in the energy convergence, indicating that the first-order ansatz already captures the dominant variational directions for this system. In contrast, for $\mathrm{BeH}_2$, the inclusion of second-order AGP-generated operators leads to a modest improvement in convergence, showing that higher-order variational directions can become useful in more demanding cases. This improvement comes with a larger operator space and additional variational parameters. For $\mathrm{BeH}_2$ at the equilibrium geometry $R_b=1.33~\text{\AA}$ considered in Fig.~\ref{fig: vqe_qaoa}(b), the number of generators increases from $N_{\mathrm{ops}}^{(l=1)}=30$ to $N_{\mathrm{ops}}^{(l=2)}=48$.

These results show that the benefit of increasing the nested-commutator truncation order is system dependent. For the molecular examples considered here, the first-order $\mbox{AGA}$ provides a practical compromise between variational flexibility and implementation cost, while higher-order expansions can become advantageous when the additional variational directions compensate for the enlarged operator space.

\section{\label{Appendix_C:} Reference ans\"atze for VQE}

In this Appendix, we briefly summarize the reference ans\"atze used for comparison with the proposed $\mbox{AGA}$ and $\mbox{AGAR}$ constructions. The 3-UpCCGSD and ADAPT-Givens implementations are carried out using PennyLane~\cite{pennylanecit,AddaptPenny}.

\subsection{UCCSD}

The unitary coupled-cluster ansatz with single and double excitations (UCCSD) is defined from the excitation operators
\begin{eqnarray}
\label{excitation_op1}
\hat{T}_1 &=& \sum_{i,a} t_i^a \hat{a}_a^\dagger \hat{a}_i,\\
\label{excitation_op2}
\hat{T}_2 &=& \frac{1}{4}\sum_{i,j,a,b}
t_{ij}^{ab}
\hat{a}_a^\dagger\hat{a}_b^\dagger
\hat{a}_i\hat{a}_j,
\end{eqnarray}
where $i,j$ label occupied orbitals and $a,b$ unoccupied orbitals. The amplitudes $t_i^a$ and $t_{ij}^{ab}$ are treated as variational parameters. Defining $\hat{T}=\hat{T}_1+\hat{T}_2$, the formal unitary ansatz is
\begin{equation}
\hat{U}_{\mathrm{UCCSD}}(\vec{\theta})
=
\exp\!\left[
\hat{T}(\vec{\theta})
-
\hat{T}^{\dagger}(\vec{\theta})
\right].
\end{equation}
The anti-Hermitian generator makes the ansatz unitary. In practice, implementation on a gate-based quantum processor requires a suitable decomposition, typically in Trotterized or disentangled form.

\subsection{$k$-UpCCGSD}

The $k$-unitary pair coupled-cluster generalized singles and doubles ($k$-UpCCGSD) ansatz reduces the excitation space by combining generalized single excitations with paired double excitations~\cite{KupCCGSD}. The corresponding excitation operators are
\begin{eqnarray}
\hat{T}_1 &=& \sum_{p,q} t_p^q
\hat{a}_p^\dagger\hat{a}_q,\\
\hat{T}_2 &=& \sum_{i,a}
t_{i_\alpha i_\beta}^{a_\alpha a_\beta}
\hat{a}_{a_\alpha}^\dagger
\hat{a}_{a_\beta}^\dagger
\hat{a}_{i_\alpha}
\hat{a}_{i_\beta},
\end{eqnarray}
where the generalized singles connect arbitrary orbitals, while the paired doubles transfer electron pairs between spatial orbitals. The $k$-layer ansatz is written as
\begin{equation}
\hat{U}_{k\text{-UpCCGSD}}
=
\prod_{j=1}^{k}
\exp\!\left[
\hat{T}^{(j)}(\vec{\theta}^{(j)})
-
\hat{T}^{(j)\dagger}(\vec{\theta}^{(j)})
\right],
\end{equation}
where each repetition has its own set of variational parameters. In this work, we use $k=3$.

\subsection{ADAPT-Givens}

ADAPT-VQE \cite{grimsley2019,AddaptPenny} is an adaptive variational protocol rather than a fixed ansatz. It iteratively constructs the ansatz by selecting operators from a predefined pool based on their contribution to the energy gradient. At iteration $n$, the gradient with respect to operator $\hat{A}_k$ is given by
\begin{equation}
\label{adapt_derivative}
    \frac{\partial E^{(n)}}{\partial \theta_k} 
    = \langle \Psi^{(n)} | [\hat{H}, \hat{A}_k] | \Psi^{(n)} \rangle.
\end{equation}
The operator with the largest gradient magnitude is added to the ansatz, and the procedure is repeated until the gradient norm falls below a threshold $\epsilon$.

In the ADAPT-Givens implementation considered here, the operator pool is constructed from particle-number-conserving Givens rotations. In the single-excitation subspace, their action can be represented as
\begin{equation}
\begin{aligned}
U_{\mathrm{GR}}(\theta)|01\rangle
&=
\cos\theta\,|01\rangle
-
\sin\theta\,|10\rangle,\\
U_{\mathrm{GR}}(\theta)|10\rangle
&=
\sin\theta\,|01\rangle
+
\cos\theta\,|10\rangle.
\end{aligned}
\end{equation}
These rotations preserve particle number and provide the operator pool used by the adaptive optimization in our benchmark calculations.

\end{document}